\documentclass[twocolumn,amsmath,amssymb]{revtex4-1}
\usepackage{dcolumn}
\usepackage{epsfig}
\usepackage{amssymb,amsmath,bm,graphicx,multirow,mathtools}
\usepackage{color}
\usepackage[colorlinks,allcolors=blue]{hyperref}
\usepackage{fleqn}
\newcommand{\cev}[1]{\reflectbox{\ensuremath{\vec{\reflectbox{\ensuremath{#1}}}}}}

\def\be{\begin{equation}}
\def\ee{\end{equation}}
\def\bea{\begin{eqnarray}}
\def\eea{\end{eqnarray}}
\def\e{\epsilon}

\def\11{1\hspace{-0.53 em}1}
\def\m{\mu}
\def\n{\nu}

\def\p{\partial}
\def\a{\alpha}
\def\b{\beta}
\def\t{\theta}
\def\d{{\rm d}}
\def\i{\imath}

\def\00{0\hspace{-0.55em}0}

\def\ma{\mathtt{A}\hspace{-0.55em}\mathtt{A}}
\def\mb{\mathtt{B}\hspace{-0.55em}\mathtt{B}}
\def\mc{\mathtt{C}\hspace{-0.58em}\mathtt{C}}

\def\hg{\hat{\g}}

\def\l{\lambda}
\def\k{\kappa}
\def\z{\zeta}

\def\hg{\hat{\Gamma}}
\def\hr{\hat{R}}

\def\d{\Delta}

\def\Gh{\hat{G}}

\begin{document}

\title{Unification of the Gauge Theories
}
\author{Abolfazl\ Jafari\footnote{1}}\email{jafari-ab@sku.ac.ir}
\affiliation{Department of Physics, Faculty of Science, Shahrekord University, Shahrekord, Iran}
\date{\today }


\begin{abstract}
\begin{center}
\center{\textbf{Abstract}}
\end{center} 

We treat the Christoffel coefficients as operators and introduce new mappings for quaternionic products to connect with the theory of electrodynamics in general spacetime.
By utilizing the directional operator of the covariant derivative, we generalize the quaternionic mechanism to the theory of electrodynamics.
We demonstrate that the Einstein equation permits the selection of a constant term that is consistent with the covariant derivative.

\end{abstract}


\maketitle


\section{Introduction}

Many attempts were made to develop a geometric approach to electrodynamics when general relativity successfully explained the universe. 
Unfortunately, most of these efforts were unsuccessful
\cite{overduin, ruiz, wes, cabral, goenner, lobo, wess, gross, ponce, williams, wesso}.
Our current goal is to pursue this objective in a significantly different manner. 
We aim to incorporate the theory of gravity into the framework of electrodynamics. 
This integration may enhance our understanding of quantum gravity, a concept that physicists widely seek to realize.
To validate the results of this work, we can analyze the components of Riemann's curvature tensor.
The quest for a deeper understanding of quantum gravity has prompted us to propose a modification in the dimensions of spacetime \cite{herbert, kiefer, gianluca, gianlu, gianluc, carlo, ashteka, ashtekar}.
In addressing this challenge, we encounter local issues related to quaternionic production, particularly regarding the contraction of indices in tangent and cotangent spaces. 
However, the fundamental outcome remains consistent, regardless of the chosen space.
These new perspectives will be instrumental in advancing our understanding of gravity theory.
We examine two structures: the spacetime metric and the quaternionic metric, which involves quaternionic production under a new metric.
As will be demonstrated, it is possible to transition from electrodynamics to gravity using quaternions. However, there is a fundamental difference between the two theories.
Electrodynamics is a comprehensive theory involving two components, described by the strength tensor in four-dimensional space.
When extending electrodynamics to a theory of gravity using quaternions, we must consider four components in a four-dimensional spacetime.
This leads to the idea of component separation.
The alteration in the structure of these components results in equations of motion, ultimately culminating in Bianchi's identity \cite{weber, misner, wald, weinberg, invarno, malcolm}.
We will show that space (geometry) has a physical entity.
Therefore, equations of motion will have to be determined, which go beyond mathematical relations to the properties of the geometry of space.
When space takes on a physical meaning, it must obey the laws of physics, and "\textit{space characters behavior must be affected by the equations of motion}".
One consequence is that Einstein's equation for gravity is merely one of the equations that determine geometry.
\newline
The outcome of this article could be a modified gravitational theory, which holds the Einstein equation by formally including additional conditions \cite{gray, parisi, faisal, sang}.
The additional conditions added to Einstein's equation are viewed strictly from a purely theoretical perspective, which geometry alone has never achieved. We will derive Einstein's equation with these additional conditions and demonstrate their necessity.
The interpretation of Einstein's historical theorem represents one of the transformative consequences of his theory. 
Understanding how gauge fields operate within the framework of gravity is a significant achievement \cite{misner, wald, invarno}. 
Inducing the behavior of gauge fields in gravity is an achievement.
The last theorem is made more complex by Riemann's four-component curvature tensor.
Certainly, the theory of gravity possesses a gauge-theoretic property analogous to gauge field theory.
The directional covariant derivative operator is a key innovation in this work.
We will discuss its effects in a subsection, but currently, we will utilize only its results.
\newline
\textbf{Clarification-}
It is essential to clarify definitions and designations:
the vector space is represented by $\varepsilon$ and its dual vector space is denoted by $\varepsilon^\star$.
They are ordered and denoted as $\varepsilon^\star \otimes \varepsilon$, and
other arrangements are not permitted.
The notation 
$\mathsf{u}^{\alpha\beta\cdots}_{\mu\nu\cdots} = u^{\alpha\beta\cdots}_{\mu\nu\cdots} \, (\hat{e}^\mu \hat{e}^\nu \cdots) \otimes (\hat{e}_\alpha \hat{e}_\beta \cdots)$ indicates a multicomponent member.
The notation is revised as follows: the components $\mathsf{u}^{\alpha\beta}_{\mu\nu} $ will now be expressed as $ u^{\alpha\beta}_{\mu\nu} $.
The interaction of indices occurs only from a vector space to the dual space and vice versa; in any case, it is from a vector space into the dual space.
\newline
\textbf{Establishing-} $"\mathsf{ast}"$ is the new mapping introduced by defining its result as the only direction-independent action of the vector on its relevant dual space,
$$\ast:\ \varepsilon\vec{\times}\varepsilon^\star\sim\mathrm{is\ deffined}$$
Also, the definition of the mapping as an action is:
$$\ast:\ \varepsilon^\star\cev{\times}\varepsilon\sim\mathrm{is\ deffined}$$
In this way,
$\varepsilon^\star\vec{\times}\varepsilon$ and $\varepsilon\cev{\times}\varepsilon^\star$
lead to invalid results.
We can also represent this with $e_\a\vec{\times} e^\b$ and so on.
We introduce a Gamma operator that is directional:
$$
\hg\ast\varepsilon^\star\rightarrow\varepsilon^\star,\ \ \ \ \varepsilon\ast\hg\rightarrow\varepsilon,
$$
Here, we will not mention the direction of the action.
Therefore, operator $\hg$ obtains a structure of $\varepsilon^\star\times\varepsilon$ 
and consequently $\hg\propto\varepsilon^\star\times\varepsilon$.
We point out that the operation of $\hg$ is the contraction.
In summary, $\hg$ is an operator in the space of indices.
\newline
Due to assumption 1, the contraction of the indices occurs up to the allowed cases; $\Sigma_{\a}\ \hat{e}_\a\ast\ (\hat{e}^\m \hat{e}^\n)=\Sigma_{\a}\ (\delta^\m_\a\hat{e}^\n+\hat{e}^\m \delta^\n_\a) $.
We assign a Lorentz index of $\hg_\m$ to the operator.
\newline
\textbf{Metric-}
According to the definition of $"\mathsf{ast}"$ mapping, the metric is a mapping of the form; 
\begin{align}\label{m1}
\mathsf{g}^{\mathrm{up}}:\varepsilon^2\times \varepsilon^\star\rightarrow\varepsilon
\end{align}
and
\begin{align}\label{m2}
\mathsf{g}_{_{\mathrm{down}}}:\varepsilon^\star\leftarrow\varepsilon\times\varepsilon^{\star 2}
\end{align}
The symbol $"\mathbf{g}"$ is the general metric element and comprises $\varepsilon\times\varepsilon$ or $\varepsilon^\star\times\varepsilon^\star$.
Equations \ref{m1} and \ref{m2} show that \textit{the metric (with a repeated index) does not obey the rules of the operator}, and it is clear that any other interpretation of equations \ref{m1} and \ref{m2} would be wrong for the action of the metric.
It is demonstrated that the metric effect does not influence the calculation of the components.
Following the definition of $"\mathsf{ast}"$ mapping, the operator $\hg_\k$ is represented as follows;
\begin{align}\label{dagger}
\hg_\k\equiv\Sigma_{\a\b}(\Gamma^\a_{\b \k}\ e^\b e_\a)^\dagger=\Gamma^{\a}_{\b \k}\ e^{\a} e_{\b}
\end{align}
We have considered the inverted version of $\hg_\k$ as the operator due to the compatibility of the Christoffel connections with the operator's role.
Furthermore, except for the $"\k"$ index in $\hg_\k$, its other two indices are repeated and dummy.
According to assumption 2, $\hg_\k\equiv\hg_\k\ e^\k$ (without summation) completes the missing part of relation equation \ref{dagger}.
From this follows (a connection coefficient of the first kind): 
\begin{align*}
\Gamma^\a_{\b\k}\mathsf{g}_{_{\mathrm{down}}}\rightarrow\Gamma^{\m}_{\b\k}\ g_{\m\n}\equiv\Gamma_{\b\k\n},
\end{align*}
the mapping is, of course, not an operator.
Due to the specificity of contracting indices, we will remove the indicator from the top of the letters.
With the notation of
\begin{align*}
e_\a\rightarrow\ <\a|
\cr
e^\a\rightarrow\ |\a>
\end{align*}
The above statements and results become easier to understand with the new representation because the $\hg_\z$ operator yields 
$"\hg_\k=\Sigma_{\a,\b}\Gamma^\a_{\b\k}|\a><\b|\equiv\Gamma^\a_{\b\k}|\a><\b|"$.
The $"\mathsf{ast}"$ mapping becomes more formal when referring to the bracket representations.
Moreover, the representation corresponding to the bracket extends our possibilities for future calculations.
The $\a$ and $\b$ are dummy indices in the $\Gamma^\a_{\b\k}|\a><\b|$-operator.
They are used, among other things, for the gamma operator effect;
\begin{align}\label{tar1}
\hg_\k\ast\mathsf{u}_\l&=\Gamma^\a_{\b \k}\ |\a><\b|\l> u_\l=\Gamma^\a_{\b \k}\ \delta_{\b\l}\ u_{_{\mathrm{vacancy}}}|\a>\nonumber\\
&=\Gamma^\a_{\l\k}u_\a,
\end{align}
During this process, the box is aligned with the alpha vector.
That is, the box takes the alpha index.
Hence,
\begin{align}\label{tar2}
\mathsf{u}^\l\ast\hg_\k=u^\l\Gamma^\a_{\b\k}\ <\l|\a><\b|=\Gamma^\l_{\b\k}u^\b.
\end{align}
We emphasize that in changing the $\mathsf{u}_\l$ notation, the position of the index $\l$ in $"u_\l"$ is the main, but the name of the index changes in 
$\ast$-multiplication.
However, we should use the $"\mathsf{ast}"$ mapping relations as simply as possible.
\newline
The relationships between the quaternion generators $\hat{q}_1=I$, $\hat{q}_2=J$, and $\hat{q}_3=K$ are defined within a Euclidean metric and adhere to the following properties: $\hat{q}_i\cdot\hat{q}_j=\delta_{ij}$ and $\hat{q}_i\times\hat{q}_j=\Sigma_k\e_{ijk}\hat{q}_k$.
If we include $\hat{q}_0=\Im$ (as the identity), we obtain a quaternion represented by the Lorentz quantities (parameters of the Lorentz index).
These are coupled and are termed “$q$ vectors.”
Lorentz quantities, which carry the Lorentz index, are represented using quaternions.
We have two representations of the Lorentz index: the four-vector and the q-vector.
We will use the notation $\mathtt{A},\ \mathtt{B},\cdots$ to denote the quaternions.
Quaternions are typically constructed from four vectors, denoted as $ A_\mu = (A_0, \mathbf{A}) $.
They can be expressed as $ \mathtt{A} = i A_0 \Im + A_k \hat{q}^k \equiv (\i \mathtt{A}_0, \mathtt{A})_q $, where the $ q $-variable signifies the expression $ (\i \mathtt{A}_0, \mathtt{A}_k \hat{q}^k) $.
Bold letters denote the spatial components of the four vectors.
The four-vector framework is based on the spacetime metric, while the Euclidean metric pertains to the quaternion representation.
The configuration significantly affects quaternionic products.
We also analyze the Lorentz index, which connects physical parameters to the quaternion generator.
We can generalize the range of coupled quantities from vectors to tensors.
For the $q$-vectors $\hg$ and $\mathtt{C}$, the extended Grassman quaternion multiplication $"\triangleright"$ (with $\ast$-product), is defined as follows \cite{andre, walker, kansu, stefano}
\begin{align}\label{circ13}
\hg\triangleright\mathtt{C}=(\hg_0\ast\mathtt{C}_0-\hg\ast\mc,\hg\vee\mathtt{C}+\hg\wedge\mc)_q,
\end{align} 
where 
$\wedge$ and $\vee$ are the vector products 
under 
the $"\mathsf{ast}"$ mapping:
\begin{align*}
\hg\wedge\mc=\e^{ijk}(\hg_i\ast\mathtt{C}_j)\hat{q}_k.
\end{align*}
and
\begin{align*}
(\hg\vee\mathtt{C})_k=\hg_0\ast\mathtt{C}_k-\hg_k\ast\mathtt{C}_0.
\end{align*}
According to the above, $\vee$ and $\wedge$ add the spatial and temporal components to the above two commutators;
\newline
i)\ The commutator associated with the $\ast$-product is represented by the symbol $[\hg,\mathtt{C}]_\ast=\hg\ast\mathtt{C}-\mathtt{C}\ast\hg$.
\newline
ii)\ Moreover, we can introduce a new commutator, i.e. the commutator associated with the multiplication of the quaternion $[\mathtt{B},\mathtt{C}]_\triangleright=\mathtt{B}\triangleright\mathtt{C}-\mathtt{C}\triangleright\mathtt{B}$ containing the product $"\ast"$, when at least one of them is operator.
There are two distinct definitions of commutators.
\newline
Based on the operator rule for the connection coefficients $\hg$, we introduce a new covariant derivative, 
\begin{align}\label{code}
\Delta_\m=\p_\m\Im-e\hg_\m
\end{align}
where 
$"e"$ is the coupling coefficient constant with the value of one.
As mentioned earlier, $\hg_\m$ contains two coupling indices.
Due to the operators $\hg$'s and the $\ast$-product $\Delta_\m$
is a semi-complete operator.
Equation \ref{code} with the $\ast$-product has the expected effects 
\begin{align*}
\Delta_\m\ast\mathtt{B}^\l_\k&=\p_\m\mathtt{B}^\l_\k
-e\Gamma^\a_{\m\k}\mathtt{B}^\l_\a\nonumber\\
&=:\mathtt{B}^\l_{\k,\m}
\end{align*}
The multiplications used in this paper refer to the $\ast$-product.
Accordingly, the effect described above dictates the following action,
\begin{align}\label{baseone2}
[\Delta_\m,(\mathtt{B}\mathtt{C})]_\ast&=\mathtt{B}[\Delta_\m,\mathtt{C}]_\ast+[\Delta_\m,\mathtt{B}]_\ast\mathtt{C}
\end{align}
It follows that we have
\begin{align}\label{mkama}
[\Delta_\m,\mathtt{B}]_\ast
&=\p_\m\mathtt{B}-e[\hg_\m,\mathtt{B}]_\ast\nonumber\\
&\equiv\mathtt{B}_{ ;\m}
\end{align}
\textbf{Quaternions-}
If we define $g^{\a\b}\hg_\b$ directly, we can write
\begin{align*}
g^{\a\b}\Delta_\b\ast\mathtt{B}_\l=g^{\a\b}\mathtt{B}_{\l,\b}=g^{\a\b}\mathtt{B}_{\l;\b}\equiv\mathtt{B}_{\l}^{\ ;\a}
\end{align*}
One can observe that $g^{\a\b}_{\ \ ;\b}=0$, and $\mathtt{B}^{\l}_{\ ;\l}=\mathtt{B}_{\l}^{\ ;\l}$.
The covariant derivative is not complete.
The incompleteness of the derivative operator is evident from equations \ref{tar1} and \ref{tar2}.
\newline
$\m$ and $\z$ are free indices in $\hg_\m\ast\mathtt{B}_\zeta$.
This means that $\hg_\m\ast\mathtt{B}_\zeta$ ends up being a second rank tensor;
In other words, $\hg_\m\ast\mathtt{B}_\zeta\propto\Gamma^\a_{\m\zeta} B_\a$ actually has the two free indices $\m$ and $\z$ that can be considered as free indices; $\Gamma^\a_{\m\zeta}B_\a\equiv\Gamma^\a_{\m\zeta}B_\a\ |\m,\zeta>$.
Equation \ref{dagger}, and our calculations in the bracket notation show that,
\begin{align*}
\hg_\n\ast\hg_\m&=\Gamma^\a_{\n\b}|\b><\a|\Gamma^\k_{\m\l}|\l><\k|+\Gamma^\a_{\n\b}|\a><\b|\m>
\nonumber\\&
\ \ \ \ \times\Gamma^\k_{\m\l}|\l><\k|
\nonumber\\&
=\Gamma^\a_{\n\b}\Gamma^\k_{\m\l}|\b>(\delta_\a^\l)<\k|+\Gamma^\a_{\n\b}|\a>(\delta_{\b\m})
\nonumber\\&
\ \ \ \ \times\Gamma^\k_{\m\l}|\l><\k|
\nonumber\\&
=\Gamma^\a_{\n\b}\Gamma^\k_{\m\a}|\b><\k|+\Gamma^\a_{\n\m}\Gamma^\k_{\a\l}|\l><\k|,
\end{align*}
furthermore 
\begin{align*}
(\hg_\n\ast\hg_\m)^\dagger&=(\Gamma^\a_{\n\b}\Gamma^\k_{\m\a}+\Gamma^\a_{\n\m}\Gamma^\k_{\a\b})|\b><\k|)^\dagger\nonumber\\
&=(\Gamma^\a_{\n\b}\Gamma^\k_{\m\a}+\Gamma^\a_{\n\m}\Gamma^\k_{\a\b})|\k><\b|,
\end{align*}
so the inverted version becomes
\begin{align*}
(\hg_\n\ast\hg_\m)\ast\mathtt{B}_\zeta=\Gamma^\a_{\n\zeta}\Gamma^\k_{\m\a}B_\k+\Gamma^\a_{\n\m}\Gamma^\k_{\a\z}B_\k.
\end{align*}
It also has the free indices, $\m$, $\n$ and $\z$.
Also,
\begin{align}\label{zarbdo}
\hg_\n\ast(\hg_\m\ast\mathtt{B}_\zeta)&=\Gamma^\a_{\n\b}|\a><\b|\ast(|\m,\zeta>\Gamma^\k_{(\m\zeta)}B_\k)\nonumber\\
&=\Gamma^\a_{\n\zeta}\Gamma^\k_{\m\a}B_\k+\Gamma^\a_{\n\m}\Gamma^\k_{\a\z}B_\k.
\end{align}
In addition,
\begin{align*}
\hg_\k\ast(\hg_\l\ast\mathtt{B}_\m)=(\hg_\k\ast\hg_\l)\ast\mathtt{B}_\m.
\end{align*}
confirms that $\Delta_\m$ is an incomplete operator, satisfying the condition:
\begin{align*}
\Delta_\m\ast(\mathtt{B}_\l\mathtt{C}_\k)=(\Delta_\m\ast\mathtt{B}_\l)\mathtt{C}_\k+\mathtt{B}_\l(\Delta_\m\ast\mathtt{C}_k).
\end{align*}
In particular, 
$$[\hg_\m,\hg_\n]_\ast=(\Gamma^\zeta_{\m\l}\Gamma^\k_{\n\zeta}-\Gamma^\zeta_{\n\l}\Gamma^\k_{\m\zeta})(|\k><\l|).$$
The free indices following the transposition are visible.
In this way, $(\Gamma^\zeta_{\m\l}\Gamma^\k_{\n\zeta}-\Gamma^\zeta_{\n\l}\Gamma^\k_{\m\zeta})(|\k><\l|)$ as an operator contains the two free indices $\k$ and $\l$.
It becomes a simplicial $[\hg_\m,\hg_\n]^\k_{\ast\ \l}$ which has free $\l$ and $\k$ indices.
Next, we follow the computation of $[\Delta_\a,\Delta_\b]_\ast$.
We compute the case without torsion,
\begin{align*}
\hg_{\kappa;\a;\b}
&=\p_\b\p_\a\hg_\kappa-e\p_\b(\hg_\a\ast\hg_\kappa)-e\Gamma^\m_{\b\a}\p_\m\hg_\kappa\nonumber\\
&-e\hg_\b\p_\a\ast\hg_\kappa+e^2\hg_\b\ast(\hg_\a\ast\hg_\kappa),
\end{align*}
also
\begin{align*}
\hg_{\kappa;\b;\a}
&=\p_\a\p_\b\hg_\kappa-e\p_\a(\hg_\b\ast\hg_\kappa)-e\Gamma^\m_{\a\b}\p_\m\hg_\kappa\nonumber\\
&-e\hg_\a\p_\b\ast\hg_\kappa+e^2\hg_\a\ast(\hg_\b\ast\hg_\kappa),
\end{align*}
consequently
\begin{align}\label{cch}
\hg_{\kappa;\a;\b}-\hg_{\kappa;\b;\a}=e\hr_{\a\b}\ast\hg_\kappa.
\end{align}
Eq.\ref{cch} means
\begin{align}\label{cch2}
[\Delta_\a,\Delta_\b]_\ast=e\hr_{\a\b}.
\end{align}
We are in a situation where the four components of Riemann's curvature tensor have merged into a single operator with two distinct subspaces.
Then, we will look at the roles of the components.
With the action role and equation \ref{zarbdo}, and for space with torsion, the new covariant derivative satisfies the following commutation relation,
$[\Delta_\a,\Delta_\b]_\ast=e\hr_{\a\b}+e^2(\Gamma^\l_{\a\b}-\Gamma^\l_{\b\a})\hg_{\l}$.
But the effect of $\hr_{\a\b}$ on variable $\mathtt{B}_\m$ is now,
\begin{align}\label{rtona}
\hr_{\a\b}\ast\mathtt{B}_\m&=(\p_\a\hg_{\b}-\p_\b\hg_{\a}-e\hg_{\a}\ast\hg_{\b}+e\hg_{\b}\ast\hg_{\a})\ast\mathtt{B}_\m\nonumber\\
&=R^\l_{\ \m\a\b}B_\l|\m,\a,\b>.
\end{align}
Another result is accessible; $[\Delta_\m,\mathsf{A}]_\ast$ reach to 
\begin{align*}
[\Delta_\m,A^\a_\b]_\ast=\p_\m A^\a_\b-g\Gamma^\l_{\b\m}A^\a_\l+g\Gamma^\a_{\l\m}A^\l_\b,
\end{align*}
and 
\begin{align*}
[\Delta_\m,\mathsf{A}]^\a_{\ast\ \b}&=(\p_\m\mathsf{A}-g\hg_\m\ast\mathsf{A}+g\mathsf{A}\ast\hg_\m)^{\a}_{\b}\nonumber\\
&=\p_\m A^\a_\b-g\Gamma^\k_{\m\z_i}A^{\cdots\z_j\cdots}_{..\z_{i-1}\k\z_{i+1}..}<\b|\vec{\z}_i><\vec{\z}_j|\a>\nonumber\\
&-A_{\cdots\z_i\cdots}^{..\z_{j-1}\k\z_{j+1}..}\Gamma^{\z_i}_{\k\m}<\b|\vec{\z}_i><\vec{\z}_j|\a>.
\end{align*}
Now, considering relations \ref{mkama} and \ref{zarbdo}, we can write:
\newline
\textbf{lemma}- The covariance derivative effect is independent of entering the components.
\newline
Since $[\hg_\m,\mathtt{A}_\n]_\ast+[\mathtt{A}_\m,\hg_\n]_\ast=
0$
, we take 
$\mathtt{A}$ to be as the gauge fields, then the change of the tensor of the field strength is
\begin{align}\label{ne}
\mathcal{E}_{\m\n}&=\p_\m\mathtt{A}_\n-\p_\n\mathtt{A}_\m\nonumber\\
&=\Delta_\m\ast\mathtt{A}_\n-\Delta_\n\ast\mathtt{A}_\m
\end{align}

\section{Electrodynamics}
In this section, we present a mechanism by which the components of gauge fields are coupled with the generators of quaternions ($\mathbf{A}\rightarrow\mathtt{A}$).
A connection has been established between the components of the covariant derivative and the quaternion generators.
The quaternion variables obtained in this way help us to reconstruct the strength tensor of the fields. Due to the semi-completeness of the covariant derivative operator, we will introduce a new quaternion commutator to obtain the equation of motion and conclude the theory.
\newline
Based on the information provided, we will define new, commonly used variables and operators for quatrains.
Starting from the tangent spaces, we introduce the operators and vectors as $q$-variables, $z_t=(\i\mathtt{A}^0,\mathtt{A}^i\hat{q}_i)$ and $D_t=(\i\Delta_0,\mathbf{\Delta}_i\hat{q}^i)$.
The position of the indices is important here. 
$\mathtt{A}$ vectors are the fields; $\ma$ is the vector potential and $\mathtt{A}_0=\phi$ represents the scalar case.
There are two more i.e. $z_c=(\i\mathtt{A}_0,\mathtt{A}_i\hat{q}^i)$ and $D_c=(\i\Delta^0,\mathbf{\Delta}^i\hat{q}_i)$.
\newline
We show that Maxwell’s equations are accessible with augmented quaternions.
The given content confirms the above definitions of the $q$-vectors.
We emphasize that the quaternion mapping creates two separate spaces: the tangent and the cotangent $q$-vectors.
The measures of the spaces are related by the new matrix elements, rather than by the Minkowski metric.
Our calculations show that the law of direct and crossed multiplication between $q$-vectors is given by
$H^{\m\n}=\left(\begin{array}{cccc}
1 & \i & \i & \i\\
-\i & -1 & 1 & -1\\
-\i & -1 & -1 & 1\\
-\i & 1 & -1 & -1\\
\end{array}\right)$, where 
\begin{align*}
H^{\m\n}=\eta^{\m\n}+\left(\begin{array}{cccc}
0 & \i & \i & \i\\
-\i & 0 & 1 & -1\\
-\i & -1 & 0 & 1\\
-\i & 1 & -1 & 0\\
\end{array}\right):=\eta^{\m\n}+H^{\m\n}_d, 
\end{align*} 
$\eta^{\m\n}=\mathrm{diag}(1,-1,-1,-1)$ is the Minkowski metric.
With the help of the matrix $H^{\m\n}$ we define the inverse matrix
\begin{align*}
\tilde{H}^{\m\n}=(H^{\m\n})^T=\left(\begin{array}{cccc}
1 & -\i & -\i & -\i\\
\i & -1 & -1 & 1\\
\i & 1 & -1 & -1\\
\i & -1 & 1 & -1\\
\end{array}\right)=\eta^{\m\n}+\tilde{H}^{\m\n}_d,
\end{align*}
\newline
We also introduce a new tensor, denoted as $n_k^{\ ij}=g_{ks}\e^{sij}$, derived from the behavior of quaternions.
This tensor has the following properties: it is antisymmetric concerning the exchange of indices and can be employed to raise and lower indices; it is spatial, although it does not involve counting, and it includes a temporal index.
Symbole $n_k^{\ ij}$ is used when its indices are on the same row.
When all indices are in a row, it resembles the Levi-Civita symbol.
We will identify the location of the temporal index later.
\newline
Since the product of the quaternions is a quaternion,
the new q-variable in the quaternion space can be named as $\mathsf{I}$, so
we obtain $D_t\triangleright\ z_c=(\i\mathsf{I}_0,\mathsf{I}_i\hat{q}^i)\equiv\mathsf{I}_c$.
Considering the indices, we obtain
$$\mathsf{I}_c=(\i\mathsf{I}_0,\mathsf{I})$$
(using the upper indices), our calculations show that
\begin{align*}
\mathsf{I}_c
&=(\mathtt{A}_{\n,\m}\ \eta^{\m\n},\mathtt{A}_{\n,\m}\ H_d^{\m\n})\nonumber\\
\end{align*}
based on this result and Eq.\ref{mkama}, we arrive at
\begin{align}\label{nq}
\i\mathsf{I}_0&=
\mathtt{A}_{\n,\m}\ \eta^{\m\n}
\end{align}
and
\begin{align*}
\mathsf{I}_k&=
\mathtt{A}_{\n,\m}\ H_d^{\m\n}
\end{align*}
are available with the following definitions:
\begin{align*}
\mathsf{I}_k&=\i \mathcal{E}_{0k}+\frac{1}{2}n^{\ \ ij}_{k}\mathcal{E}_{ij}\nonumber\\
&\equiv\i \mathcal{E}_{0k}+D_k,
\end{align*}
where $\mathcal{E}_{\m\n}$ is given by Eq.\ref{ne}.
The equation of motion describes the action of the covariant derivative on quaternion variables and vice versa.
Due to the incompleteness of the derivative effects, to obtain the equation of motion, we must define a new commutator as follows:
$[\ ,\ ]_\triangleright:\ \mathsf{Q}\times\mathsf{Q}\rightarrow\mathsf{Q}$
which is the same as
$[\mathtt{B},\mathtt{C}]_\triangleright=(\eta^{\m\n}[\mathtt{B}_\m,\mathtt{C}_\n]_\ast,[\mathtt{B},\mathtt{C}]^\vee_\ast+[\mb,\mc]^\wedge_\ast)$.
Due to the electrodynamics in Minkowski spacetime, the commutator of the motion has the following form (under 
the matrix
$\tilde{H}$):
\begin{align*}
[(\i g^{0\z}\Delta_\z,g^{s\z}\mathbf{\Delta}_\z),\mathsf{I}_c]^{\tilde{H}}_\triangleright&=(g^{\m\z}\mathsf{I}_{\n;\z}\eta^{\m\n},
\mathsf{I}_{\n;\m}\ \tilde{H}_d^{\m\n})=\mathsf{J}_c,
\end{align*}
$\mathsf{J}_c=(\i\mathsf{J}_0,\mathsf{J}_i\hat{q}^i)$ is the quaternion form of current density.
So the equation of motion expands to,
\begin{align}\label{laste0}
g^{\m\z}\mathsf{I}_{\n;\z}\eta^{\m\n}&=\i\mathsf{j}_0,\nonumber\\
\mathsf{I}_{\n;\m}\ \tilde{H}_d^{\m\n}&=\mathsf{j}_k
\end{align}
For any $\mathtt{B}$ without upper indices, the equality $\Delta\ast\mathtt{B}\equiv[\Delta,\mathtt{B}]_\ast$ holds,
and
\begin{align}\label{equality}
\Delta_\n\ast\mathtt{B}_\m=\mathtt{B}_{\m,\n}\equiv\mathtt{B}_{\m;\n}
\end{align}
If $\mathtt{B}$ carries no upper indices, then $\Delta_\n\ast\mathtt{B}_{;\m}=\mathtt{B}_{;\m;\n}$
and $\d_\m\ast g_{\k\l}=0$, $\d_\m\ast g^{\k\l}\neq0$.
Therefore, equations \ref{baseone2} and \ref{laste0} based on equation \ref{equality} lead to the following equation:
\begin{align}\label{laste01}
g^{0\z}\mathsf{I}_{0\ ;\z}-\i g^{s\z}\mathcal{E}_{0s;\z}-g^{s\z}D_{s ;\z}=\i\mathsf{J}_0,
\cr
-\i g^{0\z}(\i\mathcal{E}_{0k;\z}+D_{k ;\z})-n_{ks}^{\ \ \ l} g^{s\z}(\i\mathcal{E}_{0l;\z}+D_{l ;\z})
\cr
+\i g^{k\z}\mathsf{I}_{0\ ;\z}=\mathsf{J}_k,
\end{align}
and finally, we arrived at 
\begin{align}\label{laste1}
-\i g^{0\z}\mathsf{I}_{0\ ;\z}-g^{s\z}\mathcal{E}_{0s;\z}&=\mathsf{J}_0,\nonumber\\
g^{s\z}D_{s ;\z}&=0,\nonumber\\
\i g^{k\z}\mathsf{I}_{0\ ;\z}+g^{0\z}\mathcal{E}_{0k;\z}-n_{ks}^{\ \ \ l}g^{s\z}D_{l ;\z}&=\mathsf{J}_k,\nonumber\\
g^{0\z}D_{k ;\z}+n_{ks}^{\ \ \ l}g^{s\z}\mathcal{E}_{0l;\z}&=0
\end{align}
From the equations of motion, it follows that 
$g^{s\z}D_{s ;\z}=0$, which gives a Bianchi identity (without time index): $\frac{1}{2}n_{s}^{\ ij}g^{s\z}\mathcal{E}_{ij;\z}\rightarrow\frac{1}{2}\varepsilon^{sij}\mathcal{E}_{ij;s}=0$.
Moreover, the fourth equation is $g^{0\z}D_{k ;\z}+n_{ks}^{\ \ \ l}g^{s\z}\mathcal{E}_{0l;\z}=\frac{1}{2}n_k^{\ ij}\mathcal{E}_{ij}^{\ \ ;0}+n_{k}^{\ \ s l}\mathcal{E}_{0l;s}=0$ which gives another Bianchi identity: $\frac{1}{2} n_{k0}^{\ \ \ ij}\mathcal{E}_{ij}^{\ \ \ ;0}+n_k^{\ s0l}\mathcal{E}_{0l;s}=\frac{1}{2}\varepsilon^{k0ij}\mathcal{E}_{ij;0}+\varepsilon^{ks0l}\mathcal{E}_{0l;s}=0$.
\newline
Since $n_{ks}^{\ \ \ l}g^{s\z}D_{l ;\z}
=\frac{1}{2}\varepsilon_{ksl}g^{s\z}\varepsilon^{lij}\mathcal{E}_{ij ;\z}
=g^{s\z}\mathcal{E}_{ks ;\z}$, equation \ref{laste1} is thus
\begin{align*}
&-g^{s\z}\mathcal{E}_{0s;\z}=\mathsf{J}_0+\i g^{0\z}\mathsf{I}_{0\ ;\z}
,\nonumber\\
&g^{0\z}\mathcal{E}_{0k;\z}-g^{s\z}\mathcal{E}_{ks;\z}
=\mathsf{J}_k-\i g^{k\z}\mathsf{I}_{0\ ;\z}
\end{align*}
to summarize,
\begin{align}\label{laste2}
g^{\m\z}\mathcal{E}_{\m\n;\z}&=\mathtt{J}_\n,
\end{align}
where 
\begin{align}\label{jaraian}
\mathtt{J}_\n=\mathsf{J}_\n+\i\eta_{\n\m} g^{\m\z}\mathsf{I}_{0\ ;\z}.
\end{align}
The above relations are simplified as follows:
\begin{align*}
\mathsf{I}_0^{\ ;0}-\mathsf{I}_s^{\ ;s}=\i\mathsf{j}_0
\cr
-\i\mathsf{I}_k^{\ ;0}+\i\mathsf{I}_0^{\ ;k}-n_{ki}^{\ \ j}\mathsf{I}_j^{\ ;i}=\mathsf{j}_k
\end{align*}
From Eqs.\ref{cch} or \ref{cch2}, we get the relation $n_{ki}^{\ \ j}\mathsf{I}_j^{\ ;i;k}=\frac{e}{2}n_{ki}^{\ \ j}\hr^{ik}\ast\mathsf{I}_j$ and $\i(\mathsf{I}_k^{\ ;k;0}-\mathsf{I}_k^{\ ;0;k})=\i e\hr^{k0}\ast\mathsf{I}_k$.
If we substitute these points, we obtain 
\begin{align*}
\mathsf{j}_0^{\ ;0}+\mathsf{j}_k^{\ ;k}
&=-\i\eta_{\m\n}g^{\m\z}\mathsf{I}_{0;\z}^{\ \ ;\n}+\i e\hr^{k0}\ast\mathsf{I}_k-\frac{e}{2}n_{ki}^{\ \ j}\hr^{ik}\ast\mathsf{I}_j\nonumber\\
&=-\i\eta_{\m\n}g^{\m\z}\mathsf{I}_{0;\z}^{\ \ ;\n}\nonumber\\
&+\frac{\i e}{2}(n_{k}^{\ ij}\hr^{k0}\ast\mathcal{E}_{ij}-n_{ki}^{\ \ j}\hr^{ik}\ast\mathcal{E}_{0j})\nonumber\\
&-\frac{e}{4}(4\hr^{k0}\ast\mathcal{E}_{0k}+n_{ki}^{\ \ j}n_{j}^{\ mn}\hr^{ik}\ast\mathcal{E}_{mn})
\end{align*}
However, it can be seen that $n_{ki}^{\ \ j}n_{j}^{\ mn}\hr^{ik}\ast\mathcal{E}_{mn}=\varepsilon_{kij}\varepsilon^{jmn}\hr^{ik}\ast\mathcal{E}_{mn}=2\hr^{nm}\ast\mathcal{E}_{mn}=2\hr_{\n\m}\ast\mathcal{E}^{\m\n}-4\hr_{0\m}\ast\mathcal{E}^{\m0}$.
Moreover, the second and third terms are zero because the equations of motion
$n_{k}^{\ ij}\hr^{0k}\ast\mathcal{E}_{ij}+n_{ki}^{\ \ j}\hr^{ik}\ast\mathcal{E}_{0j}=0$.
This result is independent of the physical conditions and has nothing to do with the sources.
Substituting the results, we get 
\begin{align}\label{yeki}
\mathsf{j}_0^{\ ;0}+\mathsf{j}_k^{\ ;k}+\i\eta_{\m\n}g^{\m\z}\mathsf{I}_{0;\z}^{\ \ ;\n}=-\frac{e}{2}\hr^{\n\m}\ast\mathcal{E}_{\m\n}
\end{align}
\newline
Now we derive equation \ref{laste2} and from the fact that 
\begin{align*}
g^{\n\k}\Delta_\k\ast g^{\m\z}\mathcal{E}_{\m\n;\z}&=g^{\n\k}\Delta_\k\ast\mathtt{J}_\n,
\end{align*}
we have: $g^{\m\z}\mathcal{E}_{\m\n;\z}
=g_{\n\b}\mathcal{E}^{\a\b}_{\ \ ;\a}$
On the other hand, $g_{\n\b}\mathcal{E}^{\a\b}_{\ \ ;\a}$ and the action of the derivative operator on the $\n$ set is proportional to the low index set. From this follows: $g^{\n\k}\Delta_\k\ast g_{\n\b}\mathcal{E}^{\a\b}_{\ \ ;\a}
=\mathcal{E}^{\a\b}_{\ \ \ ;\a;\b}$.
The same logic applies to the $g^{\n\k}\Delta_\k\ast\mathtt{J}_\n=\mathtt{J}_{\n ;\k}g^{\k\n}$.
With these interpretations of equation \ref{rtona}, we can write: $\mathcal{E}^{\m\n}_{\ \ ;\m;\n}
=-\mathcal{E}^{\n\m}_{\ \ ;\n;\m}+e\hr_{\m\n}\ast\mathcal{E}^{\m\n}$,
and finally, we have
\begin{align}\label{deki}
\mathcal{E}^{\m\n}_{\ \ ;\m;\n}=\frac{e}{2}\hr_{\m\n}\ast\mathcal{E}^{\m\n}
\end{align}
\newline
According to equations \ref{yeki}, \ref{deki}, and the above, the continuity equation is as follows: 
\begin{align*}
\mathsf{j}_0^{\ ;0}+\mathsf{j}_k^{\ ;k}+\i\eta_{\m\n}g^{\m\z}\mathsf{I}_{0;\z}^{\ \ ;\n}=\frac{e}{2}\hr^{\m\n}\ast\mathcal{E}_{\m\n},
\end{align*}
But, $\mathcal{E}_{\m\n}^{\ \ ;\m;\n}=A_{\n;\m}^{\ \ ;\m;\n}-A_{\m;\n}^{\ \ ;\m;\n}=e\hr^{\m\n}A_{\n;\m}$ and based on the relation in \ref{rtona}, we reach $\mathcal{E}_{\m\n}^{\ \ ;\m;\n}\rightarrow-eR_\m^{\ \l\m\n}A_{\l;\n}-eR_\n^{\ \l\m\n}A_{\m;\l}=-eR_\n^{\ \l\n\m}\mathcal{E}_{\m\l}=0$, which means that the condition $\mathtt{J}_\n^{\ ;\n}=0$ as continuity equation is in agreement with reference \cite{lobo}.
Inserting equation \ref{ne}, equation \ref{laste2} becomes 
\begin{align*}
g^{\m\zeta}(\Delta_\m\ast A_\n-\Delta_\n\ast A_\m)_{;\z}&=\mathtt{J}_\n,
\end{align*} 
similar to the result of \cite{arbabb}, or 
\begin{align*}
g^{\m\zeta}\Delta_\zeta\ast\Delta_\m\ast A_\n-g^{\m\zeta}\Delta_\zeta\ast\Delta_\n\ast A_\m&=\mathtt{J}_\n,
\end{align*}
but 
$g^{\m\zeta} A_{\m;\n;\z}=g^{\m\zeta}(\Delta_\n\ast\Delta_\z+e\hr_{\n\z})\ast A_\m$, which gives us 
\begin{align*}
g^{\m\zeta}\Delta_\zeta\ast\Delta_\m\ast A_\n-eg^{\m\zeta}\hr_{\n\z}\ast A_\m&=\mathtt{J}_\n+
g^{\m\z}\Delta_\n\ast\Delta_\z\ast A_\m.
\end{align*}
Since $g^{\m\z}\Delta_\n\ast\Delta_\z\ast A_\m=g^{0\z}A_{0;\z;\n}+g^{s\z}A_{s;\z;\n}$ and the given condition that $A^\k_{\ ;\k}=0=g^{\k\l}A_{\l;\k}=g^{0\k}A_{0;\k}+g^{s\k}A_{s;\k}$, we can derive that $g^{s\z}A_{s;\z}=-g^{0\z}A_{0;\z}$.
We obtain the following equation \cite{parker, parke, lindgren, arbab, cameron}:  
\begin{align*}
g^{\m\zeta}\Delta_\zeta\ast\Delta_\m\ast A_\n-eg^{\m\zeta}R^\a_{\ \m\n\zeta}A_\a&=\mathtt{J}_\n,
\end{align*}
is, by substitution of the Ricci tensor, equivalent to
\begin{align}\label{feq}
g^{\m\zeta}\Delta_\zeta\ast\Delta_\m\ast A_\n-e\mathrm{R}^{\ \ \a}_{\n}A_\a&=
\mathtt{J}_\n,
\end{align}
which is
\begin{align*}
g^{\m\zeta}A_{\n;\m;\z}-e\mathrm{R}^{\ \ \a}_{\n}A_\a&=
\mathtt{J}_\n.
\end{align*}
To reach the final goal (Eq.\ref{feq}), the relation 
$\mathtt{A}^\k_{\ ;\k}=0$
must be established, and this is a precondition.
With the above gauge, the continuity equation will be: 
$\mathsf{j}^\k_{\ ;\k}=0$.
\newline
To determine the position of the time index in the new tensor $ n_k^{\ ij} $, it is sufficient to define the following:
\begin{align*}
G^{\m\n}=\frac{1}{2}\varepsilon^{\m\n\a\b}\mathcal{E}_{\a\b},
\end{align*}
so that $G^{\m\n}_{\ \ ;\n}=0$.
For the case $\m=0$ the second equation of motion in Eq.\ref{laste1} is now valid:
\begin{align}\label{ti1}
G^{0k}_{\ \ ;k}=\frac{1}{2}\varepsilon^{0sij}\mathcal{E}_{ij;s},
\end{align}
and for the case $\m\neq0$ we get the fourth equation of motion in Eq.\ref{laste1}
\begin{align}\label{ti2}
G^{l\z}_{\ \ ;\z}=\frac{1}{2}\varepsilon^{l0ij}\mathcal{E}_{ij;0}+\varepsilon^{li0j}\mathcal{E}_{0j;i}=0
\end{align}
which gives the position of the time index in the comparison.
\section{Geometry}
In this section, the gauge fields are replaced by the Christoffel connections $\mathsf{A}$ to $\Gamma$.
This change does not affect the definition of the covariant derivative operator.
Using the same method of the previous section and multiplying the four components of the Riemann’s curvature tensor into two two-component subspaces, we define the quaternion variables as $z_c=(\i\hg_0,\hg_i\hat{q}^i)$ and $D_c=(\i\Delta^0,\mathbf{\Delta}^i\hat{q}_i)$.
It is clear that $\hg^0=g^{0\z}\hg_\z$, and this is a general case.
If we set the above points and emphasize the position of the indicators, we can write 
\begin{align*}
\mathsf{I}_c
&=(\hg_{\n,\m}\eta^{\m\n},\hg_{\n,\m}\tilde{H}_d^{\m\n})
\end{align*}
With the help of these preliminary remarks, and similar to the relation in \ref{nq}, we arrive at the definitions 
\begin{align*}
\i\mathsf{I}_0&=\hg_{\n,\m}\eta^{\m\n}
\end{align*}
and the equation
\begin{align*}
\mathsf{I}_k=\hg_{\n,\m}\tilde{H}_d^{\m\n},
\end{align*}
is presented in shorthand form as follows:
\begin{align*}
\mathsf{I}_k&=\i\mathcal{F}_{0k}+\frac{1}{2}n_{k}^{\ ij}\mathcal{F}_{ij}
\nonumber\\
&=\i\mathcal{F}_{0k}+\mathfrak{D}_k
\end{align*}
The new notation $\mathcal{F}_{\m\n}$ (Eq.\ref{ne}) serves to visually distinguish this variable from that of the previous section, $\mathcal{E}_{\m\n}$.
As a result, all previous relationships and their arrangements remain intact with this transformation.
Thus, the equation of motion for the new fields $\mathcal{F}_{\m\n}$ will be found in the relation in \ref{laste01}.
By entering the quaternion current density and separating the main equations into real and imaginary components, we have
\begin{align}\label{eqm}
-\i g^{0\z}\mathsf{I}_{0\ ;\z}-g^{s\z}\mathcal{F}_{0s;\z}&=\mathsf{J}_0,\nonumber\\
g^{s\z}D_{s ;\z}&=0,\nonumber\\
\i g^{k\z}\mathsf{I}_{0\ ;\z}+g^{0\z}\mathcal{F}_{0k;\z}-n_{ks}^{\ \ \ l}g^{s\z}D_{l ;\z}&=\mathsf{J}_k,\nonumber\\
g^{0\z}D_{k ;\z}+n_{ks}^{\ \ \ l}g^{s\z}\mathcal{F}_{0l;\z}&=0
\end{align}
There is a clear connection between the sources and dynamics of geometric tensors.
Considering the results from the previous section, it is evident that the properties of the gauge field strength tensor apply to both pair components of the Riemann curvature tensor.
These performances emphasize that the Christoffel connections have the gauge property.
Equations \ref{ti1} and \ref{eqm} give the source less equations: 
\begin{align*}
\frac{1}{2}\varepsilon^{0kij}\hr_{ij;k}=0.
\end{align*}
and from equation \ref{ti2}
\begin{align}\label{002}
\frac{1}{2}\varepsilon^{l0ij}\hr_{ij;0}+\varepsilon^{li0j}\hr_{0j;i}=0
\end{align}
Eq.\ref{002} represents the Bianchi identities and can be derived from $\Gh^{\m\n}=\frac{1}{2}\varepsilon^{\m\n\a\b}\hr_{\a\b}$.
For the Bianchi identities, the sub-components of the released space behave independently.
But, in the second relation, which is independent of the source, the behavior of the second Bianichi identity is intertwined with the basis of the subcomponents.
The remaining two relations of the basic equation of motion (Eq.\ref{eqm}) are as follows:
\begin{align}\label{laste22}
g^{\m\z}\hr_{\m\n;\z}&=\mathtt{J}_\n,
\end{align}
which $\mathtt{J}_\n$ in Eq.\ref{laste22}, still gives the current density as Eq.\ref{jaraian}.
together with $\hr_{\m\n}^{\ \ ;\m;\n}=0$ and the condition $\mathtt{J}_\n^{\ ;\n}=0$.
Besides the gauge condition, the precondition $$\eta_{\n\m} g^{\m\z}\mathsf{I}_{0\ ;\z}=0,$$ is also conceivable.
\newline
A new tensor, $G^{\k\l}_{\ \ \ \m\n}$, can be introduced here: 
\begin{align*}
G^{\k\l}_{\ \ \ \m\n}=\frac{1}{4}\varepsilon^{\k\l\a\b}\varepsilon_{\m\n\t\pi}R^{\t\pi}_{\ \ \ \a\b},
\end{align*}
which gets its special form:
\begin{align*}
G^{\k\l}_{\ \ \ k\n}
\nonumber\\
&=-\frac{1}{2}\delta^\l_\n R_{\a\b}^{\ \ \ \a\b}+R_{\n\a}^{\ \ \ \l\a}.
\end{align*}
On the other hand, if we use the equation of motion (Eq.\ref{eqm})
\begin{align}\label{eeq}
G^{\k\l}_{\ \ \ \m\n;\k}&=\frac{1}{4}\varepsilon^{\k\l\t\pi}\varepsilon_{\m\n\a\b}R^{\a\b}_{\ \ \ \t\pi;\k}=0\nonumber\\
&=\frac{1}{2}\varepsilon^{\a\l\t\pi}\varepsilon_{\m\n\a\b}\mathtt{J}^{\b}_{\ \ \t\pi}-\frac{1}{2}\delta_\n^\l\ \mathcal{R}_{;\m}\nonumber\\
&
+R^{\l\a}_{\ \ \ \n\a;\m},
\end{align}
$\mathcal{R}$ is the Ricci scalar.
As far as we know
\begin{align*}
\varepsilon^{\a\l\t\pi}\varepsilon_{\m\n\a\b}\mathtt{J}^\b_{\t\pi}&=-\delta^{\a}_{\a}\det{\left(\begin{array}{ccc}
\delta^{\l}_{\m} & \delta^{\l}_{\n} & \delta^{\l}_{\b} \\
\delta^{\t}_{\m} & \delta^{\t}_{\n} & \delta^{\t}_{\b} \\
\delta^{\pi}_{\m} & \delta^{\pi}_{\n} & \delta^{\pi}_{\b} 
\end{array}\right)}\mathtt{J}^\b_{\t\pi}\nonumber\\
&=(4-1)!(-2\delta^{\l}_{\m}\mathtt{J}^\b_{\n\b}+2\delta^{\l}_{\n}\mathtt{J}^\b_{\m\b}-2\delta^{\l}_{\b}\mathtt{J}^\b_{\m\n})
\nonumber\\
&:=\mathtt{S}^\l_{\m\n},
\end{align*}
so Eq.\ref{eeq} becomes
\begin{align}\label{veq}
-\frac{1}{2}\delta^\l_\n\ \mathcal{R}_{;\m}+R^{\l\a}_{\ \ \ \n\a;\m}+\mathtt{S}^{\l}_{\m\n}=0,
\end{align}
if one now sets $\mathtt{S}^\l_{\m\n}:=\mathtt{T}^\l_{\n;\m}$ and that based on the equation \ref{mkama}, $-\frac{1}{2}\delta^\l_\n\ \mathcal{R}_{;\m}+R^{\l\a}_{\ \ \ \n\a;\m}=[\Delta_\m,\frac{1}{2}\delta^\l_\n\ \mathcal{R}+R^{\l\a}_{\ \ \ \n\a}]_\ast$, 
one finds a more complete form of \ref{veq},
\begin{align*}
-\frac{1}{2}\delta^\l_\n\ \mathcal{R}+R^{\l\a}_{\ \ \ \n\a}=-\mathtt{T}^{\l}_{\n}+\mathtt{C}^\l_{\n}
\end{align*}
certainly 
\begin{align}\label{Eec}
\mathtt{C}^{\l}_{\n;\m}=0
\end{align}
is like the famous case $\mathtt{C}_{\l\n}=-\Lambda g_{\l\n}$.
It turns out that $\mathtt{C}_{\l\n}=f_{\l\n}(\mathbf{g})$ (for example $(e^{\k\mathbf{g}})_{\l\n}$, $\mathbf{g}$ is the matrix for the metric).
$\mathtt{C}_{\l\n}$, is a geometrical quantity, not matter.
The tensor $\mathtt{T}_{\l\n}$ absorbs the general background matter influence, and the effect becomes zero.
\newline
In the case of lower indices, 
\begin{align}
-\frac{1}{2}g^{\l\b} g_{\b\n}\ \mathcal{R}+g^{\l\b}R^{\ \a}_{\b \ \n\a}=-g^{\l\b}\mathtt{T}_{\b\n}+g^{\l\b}\mathtt{C}_{\b\n}
\end{align}
or
\begin{align}\label{gr}
\mathrm{R}_{\l\n}-\frac{1}{2}g_{\l\n}\ \mathcal{R}+\mathtt{C}_{\l\n}=\mathtt{T}_{\l\n}
\end{align}
In the final equation, the characteristics of geometry and matter are distinguished.
The fact that in Einstein’s theory of gravitation, Einstein’s term (in agreement with the covariant derivative) is not the only way to choose $\mathtt{C}_{\l\n}$ is another result of this mechanism.
We must also consider the constants that indicate the limitations of spacetime when solving the equation of gravity.
It is a question of finding the constant from equation \ref{gr} because it is not a unique quantity.
Changing any principal constant in equation \ref{gr} (agreement with covariant derivative) leads us into a new universe. Among the possible options for the constant $\mathtt{C}_{\l\n}$, equation \ref{gr} makes a perturbation theorem.
What can be obtained, however, is the non-uniqueness of Einstein’s famous theorem.
\newline
\section{Conclusions}
The results of this article may provide solutions to gravitational issues such as dark matter and be forbidden on Einstein's equation to take additional terms.
It is possible to modify Einstein's equation by adding terms that have vanished covariant derivatives.
\newline
\section{Discussion}

Here, we have used a quaternion mechanism to form new vectors.
We have provided $q$-vectors as physical quantities, events in the manifold of spacetime that correspond to the position of the Lorentz indices, which have coupling with the generators of quaternion algebra.
Employing operator $\hg_\m$ and mapping $\mathsf{ast}$ ($\ast$-product), we successfully developed the theory of gauge fields in a general spacetime.
We introduce the new antisymmetric tensor $n_k^{\ ij}$ and use the equations of motion of gauge theory to obtain the Bianchi identities.
We have demonstrated that the Bianchi identities represent physical equations of motion rather than merely mathematical relationships.
As we proceeded, we replaced the gauge fields with the Christoffel coefficients and followed the improved quaternion mechanism to access the theory of gravity. 
Our calculations indicate that with this substitution and assuming more degrees of freedom for Riemann's curvature tensor, the theory of gravitation remains valid, and Einstein's constant term for the equations of motion can be formally defined.
The results of the theory of gravity are much higher than those of electrodynamic theory, and the reason is Riemann’s curvature four-component tensor.
As for the Bianchi identities in gravity, we have concluded that the identities are part of the equations of motion.
One of the most significant outcomes of deriving the equation of gravity is the formalization of the constant term, referred to as Einstein's constant term, in the Einstein equation.
We have also established that the constant term is exclusively geometric. 
Having decomposed the four-component Riemann’s curvature tensor into one with two component subsections, another result of this work is to emphasize the independent behavior of the two-component subsections of the curvature tensor.
The result of the present work is a commentary on \cite{parisi, faisal}.
Many other references fit the physical data with the solutions of Einstein's equation, along with introducing many modifications to Einstein's theory of gravity and using non-principled correction approaches to Einstein's equation (corrections that are not applied to the equation \ref{Eec}).
By analyzing the components of Riemann's curvature tensor, we develop a quaternion approach and we can suggest a new way to quantum gravity, with additional equations of motion.
This article concludes that any manipulation of Einstein's equation is wrong except to meet the condition that the covariant derivative is zero.\section{Acknowledgments}
The author would like to thank Shahrekord University for providing the research grant.
\newline
\section{References}


\begin{thebibliography}{99}



\bibitem{overduin} J. M. Overduin, P. S. Wesson, Physics Report {\bf 283}, (1997) 303-378
\bibitem{ruiz}     G. Atondo-Rubio, J.A. Nieto, L. Ruiz, and J. Silvas, Revisia Mexicana De Fizica {\bf 54}, 3 (2008) 188–193
\bibitem{wes}      P. S. Wesson, Five-dimensional physics: classical and quantum consequences of Kaluza-Klein Cosmology, (World Scientific Publishing Co. Pte. Ltd, 2006)
%
\bibitem{cabral}   F. Cabral, F. S. N. Lobo, Foundations of Physics {\bf 47}, 2 (2017) 208-228
%
\bibitem{goenner}  Hubert F. M. Goenner, Living Reviews in Relativity {\bf 7}, 2 (2004)
%
\bibitem{lobo}     F. Cabral, F. S. N. Lobo, Eur. Phys. J. Plus {\bf 132}, (2017)
%
\bibitem{wess}     P. S. Wesson, Spacetime Matter, Modern Kaluza–Klein Theory, (World Scientific Pub Co Inc. 1999)
%
\bibitem{gross}    David J. gross, M. J. Perry, Nucl. Phys. B {\bf 226}, 1 (1983) 29-48
\bibitem{ponce}    P. S. Wesson, J. Ponce de Leon, Astron. Astrophys. {\bf 294}, (1995) 1-7
\bibitem{williams} L. Williams, 48th AIAA/ASME/SAE/ASEE {\bf AIAA} (2012) 3916
%
\bibitem{wesso}    P. S. Wesson, International Journal of Modern Physics D {\bf 24}, 1 (2015) 1530001
\bibitem{herbert}  Herbert W. Hamber, Quantum Gravitation, (Berlin Heidelberg: Springer-Verlag, 2009)
\bibitem{kiefer}   C. Kiefer, Quantum Gravity, (Oxford University Press Inc. New York 2007)
\bibitem{gianluca}  G. Calcagni, Class. Quantum Grav. {\bf 38}, 16 (2021) 165005
                   and, Class. Quantum Grav. {\bf 38}, 16 (2021) 165006
%
\bibitem{gianlu}   G. Calcagni, M. Ronco, Nucl. Phys. B {\bf 923}, (2017) 144-167
\bibitem{gianluc}  G. Calcagni, D. Rodriguez Fernandez, M. Ronco, Eur. Phys. J. C. Part Fields {\bf 77}, 5 (2017) 335.
%
\bibitem{carlo}     C. Rovelli, Living Reviews in Relativity {\bf 1}, 1 (1998)
%
\bibitem{ashteka}   A. Ashtekar, Phys. Rev. Lett. {\bf 57}, 18 (1986) 2244–2247 
%
\bibitem{ashtekar} A. Ashtekar, C. Rovelli, Class. Quantum Grav. {\bf 9}, 5 (1992) 1121–1150
                   and A. Ashtekar, C. Rovelli, L. Smolin, Phys. Rev. D {\bf 44}, 6 (1991) 1740–1755

\bibitem{weber}    J. Weber, General Relativity and Gravitational waves, (New York: Dover Publications, INC. 1961)
%
\bibitem{misner}   C. W. Misner, Kip S. Throne, and John Archibal Wheeler, Gravitation, (San Fransisco: W. H. Freeman and Company, 1973)
%
\bibitem{wald}     Robert M. Wald, General Relativity, (Chicago: University of Chicago Press, 1984)
%
\bibitem{weinberg} S. Weinberg, Gravitation and Cosmology: Principles and Applications of the General Theory of Relativity,  (New York: John Wiley $\&$ Sons, Inc. 1972)
%
\bibitem{invarno}  R. D'inverno, Introducing Einstein's Relativity, (New York: Oxford University Press Inc. 1998) 
%
\bibitem{malcolm}  M. Ludvigsen, General Relativity, (Cambridge University Press, 2004) 
%
\bibitem{gray}     Gary T. Horowitz, Robert M. Wald, Phys. Rev. D {\bf 17}, 2 (1978)
%
\bibitem{parisi}   L. Parisi, R. Canonico, Journal of Geometry and Symmetry in Physics (JGSP) {\bf 22}, (2011) 51–65
%
\bibitem{faisal}   A. Y. Abdelmohssin Faisal, Progress in Physics {\bf 14}, (Printed in the United States of America: 1-4 2018)
%
\bibitem{sang}     Sangwha-Yi, International Journal of Advanced Research in Physical Science (IJARPS) {\bf 8}, 9 (2021) PP 5-7
\bibitem{andre}    A. Waser, Quaternions in electrodynamics, On-line: http://www. aw-verlag (2000)   
%
\bibitem{walker}   M. J. Walker, American Journal of Physics {\bf 24}, (1956) 515-522
%
\bibitem{kansu}    M. E. Kansu, M. Tanish, and S. Demir, Eur. Phys. J. Plus (2020) 135:187
%
\bibitem{stefano}  S. De Leoy, P. Rotelli, J. Phys. G: Nucl. Part. Phys. {\bf 22}, (1996) 1137–1150 
%
\bibitem{arbabb}   A. I. Arbab, Optik, International Journal for Light and Electron Optics, {\bf 241} (2021) 167009
\bibitem{parker}   L. Parker, Phys. Rev. Let. {\bf 44}, 23 (1980)
%
\bibitem{parke}    L. Parker, Phys. Rev. D {\bf 22}, 8 (1980)
%
\bibitem{lindgren} J. Lindgren, J. Liukkonen, J. Phys. Conf. Ser. {\bf 1956}, (2021) 012017 
%
\bibitem{arbab}    A. I. Arbab, Optik {\bf 247}, (2021) 167940 
%
\bibitem{cameron}  C. R.D. Bunney, G. Gradoni, IEEE Ant. Propagation Magazine {\bf 64}, 3  (2022)

\end{thebibliography}
\end{document}